\documentclass[
 reprint,
superscriptaddress,
 amsmath,amssymb,
 aps,
]{revtex4-2}

\usepackage{graphicx}
\usepackage{dcolumn}
\usepackage{bm}
\usepackage{orcidlink}

\usepackage[scr=boondox]{mathalfa}
\hypersetup{
    colorlinks=true,
    linkcolor=red,
    citecolor=blue,
    urlcolor=blue
}

\begin{document}

\preprint{APS/123-QED}

\title{
Improved Bounds on the Dark Matter-Neutrino Cross Section using LSBGs}

\author{Akari Suezaki}
  \email{suezaki.a.c0e1@m.isct.ac.jp}
\affiliation{Department of Physics, Institute of Science Tokyo,\\
2-12-1 Ookayama, Meguro-ku, Tokyo 152-8551, Japan}

\author{Shunsaku Horiuchi \orcidlink{0000-0001-6142-6556}}
  \email{horiuchi@phys.sci.isct.ac.jp}
  
\affiliation{Department of Physics, Institute of Science Tokyo,\\
2-12-1 Ookayama, Meguro-ku, Tokyo 152-8551, Japan}
\affiliation{Kavli IPMU (WPI), UTIAS, The University of Tokyo, Kashiwa, Chiba 277-8583, Japan}
\affiliation{Center for Neutrino Physics, Department of Physics, Virginia Tech, Blacksburg, VA 24061, USA}





\date{\today}

\begin{abstract}
    We derive new constraints on the interaction cross section between dark matter and neutrinos ($\sigma_{\nu\chi}$) that improve upon existing limits. We adopt a mechanism where neutrinos emitted from core-collapse supernovae transfer their energy to the dark matter halo via elastic scattering, transitioning a centrally cusped profile to a flat-cored profile. We apply this mechanism to Low Surface Brightness Galaxies (LSBGs), which are characterized by their relatively large sizes, low stellar densities, and measurements of dark matter cores. We improve upon previous methods so that each dark matter particle is counted at most once, which weakens existing limits, e.g., from the Fornax dSph by a factor of $\sim 50$. Applying our new method to LSBGs, we derive an upper limit of $\sigma_{\nu\chi}< 8.5 \times 10^{-25}\,\rm cm^2$ for a dark matter mass of $m_\chi =1\,\rm GeV$ at a typical supernova neutrino energy of $E_\nu=15\,\rm MeV$. This constraint is approximately three orders of magnitude stronger than the Fornax limit derived with the same method and slightly stronger than the bound from the dark matter diffused supernova neutrino background (DMDSNB).
\end{abstract}

\maketitle

\newpage
\section{Introduction}
Although dark matter is known to constitute approximately 27\% of the energy density of the Universe, its fundamental nature and interactions with Standard Model particles remain elusive \cite{Bertone:2004pz, Planck:2018vyg,Feng:2010gw}. While direct detection experiments have placed highly stringent constraints on the dark matter interactions with nucleons and electrons \cite{LZ:2022lsv,XENON:2023cxc,  PandaX-4T:2021bab, DarkSide:2018ppu, SENSEI:2020dpa, SuperCDMS:2018mne}, the interaction between dark matter and neutrinos remains poorly constrained \cite{Wilkinson:2014ksa,Akita:2023yga,Fujiwara:2023lsv,Cline:2022qld,Ferrer:2022kei,Chauhan:2025hoz}.

Since neutrinos interact only weakly with other particles, current bounds on the dark matter-neutrino interactions often rely on astrophysical and cosmological observations, which span a wide range of energy scales. In the sub-eV energy regime, the best limits are provided by cosmological observations such as the cosmic microwave background (CMB) and the Lyman-$\alpha$ forest, where neutrino-dark matter ($\nu-\rm DM$) interactions suppress primordial density fluctuations through collisional damping and dark acoustic oscillations (DAO) \cite{Mangano:2006mp, Wilkinson:2014ksa, Mosbech:2026xcc,Mosbech:2020ahp, Hooper:2021rjc}. In the high-energy regime of TeV-PeV, by considering the DM spike around the central supermassive black hole, neutrino telescopes such as IceCube present the best constraints by observing active galactic nuclei (AGNs), such as the blazar TXS 0506+056 \cite{Cline:2022qld,Ferrer:2022kei}.
In the intermediate energy range of $\sim \mathcal O(10)\,\rm MeV$, bounds are primarily derived from SN1987A, which is the only core-collapse supernova (CCSN) where a neutrino signal has been detected so far \cite{Kamiokande-II:1987idp,Alekseev:1988gp,Bionta:1987qt, Mangano:2006mp}. Among these wide-range constraints, we focus on placing stronger bounds in the intermediate range corresponding to the typical energy of CCSN neutrinos.

A promising avenue is so exploit the effects on structure formation. While the standard Cold Dark Matter (CDM) model successfully explains the large-scale structure and CMB data \cite{Planck:2018vyg, eBOSS:2020yzd, SDSS:2003eyi}, it has challenges on observations of small-scale structures \cite{Bullock:2017xww,DelPopolo:2016emo}. These issues include the missing satellites problem \cite{Klypin:1999uc, Moore:1999nt}, where simulations overestimate the number of dwarf satellite galaxies orbiting Milky Way-sized galaxies, and the too-big-to-fail problem \cite{Boylan-Kolchin:2011qkt, Papastergis:2014aba, Boylan-Kolchin:2011lmk}, where massive subhalos, which are expected to host bright satellites, are predicted to be too dense compared to observations. Another prominent example is the core-cusp problem \cite{Moore:1999gc, Moore:1994yx, Flores:1994gz}: while N-body simulations of collisionless CDM predict a cuspy density profile toward the galactic center, observations of various galaxies favor a cored central density profile.

To resolve the core-cusp problem, various mechanisms have been proposed. A leading astrophysical solution is baryonic feedback, where gas outflows from supernovae alter the gravitational potential, flattening the dark matter cusp into a core. Indeed, many such studies successfully bridge the difference between simulations and observations \cite{Mashchenko:2007jp, Governato:2009bg, Pontzen:2011ty, DiCintio:2013qxa, Chan:2015tna, Dutton:2016okm, Tollet:2015gqa, Freundlich:2019lkx, Burger:2021sep}. On the other hand, alternative approaches propose modifying the nature of dark matter itself, such as self-interacting dark matter (SIDM) \cite{Spergel:1999mh, Rocha:2012jg, Tulin:2017ara, Elbert:2014bma, Vogelsberger:2012ku, Zavala:2012us} and ultra-light axions or fuzzy dark matter (FDM) \cite{Hu:2000ke, Schive:2014dra, Marsh:2015wka, Hui:2016ltb} to form cores. 

DM--neutrino interactions have previously been considered as a remedy by forming cores \cite{Heston:2024ljf}. In this mechanism, $\sim 15 \, \rm MeV$ neutrinos emitted from past core-collapse supernovae (CCSNe) scatter off DM particles in the halo, transferring their kinetic energy. This energy injection causes the DM particles to expand outward, transforming an initial cusp into a core. By applying this mechanism to dwarf spheroidal galaxies (dSphs), they reported a constraint slightly more stringent than that derived from the flux attenuation of SN1987A \cite{Mangano:2006mp}.

In this work, we extend this framework to Low Surface Brightness Galaxies (LSBGs). LSBGs are highly dark-matter dominated, their core sizes are well determined from rotation curves, and their stellar masses are much larger than those of dSphs, implying a larger number of CCSN neutrinos that can scatter off DM particles. We also revise the method of \cite{Heston:2024ljf} so that each DM particle is not counted more than once, which weakens the Fornax dSph limit by a factor of $\sim 46$. Applying the revised method to both LSBGs and Fornax, we find that the resulting upper bounds on the cross section for LSBGs is about three orders of magnitude stronger than that of Fornax, and still stronger than the original Fornax limit by a factor of $\sim 40$.

This paper is organized as follows: In Sec. \ref{methods}, we describe our analytical framework, including the dark matter density profiles, the selection of LSBG sample, the neutrino emission from CCSN, and the derivation of the scatterign cross section. In Sec. \ref{results}, we present our main results and discuss their implications. We also make a comparison between our derived cross section at the DM mass $m_\chi =1\, \rm GeV$ and other constraints obtained from various observations. Finally in Sec. \ref{conclusion}, we conclude our study by stating the characteristics of the galaxies that provide the most stringent constraints, and referring to the underlying uncertainties in our analysis.

\section{\label{methods}Method}
In this section, we present the analytical framework and procedures to derive the upper bounds on the dark matter-neutrino scattering cross section. First, we introduce the density profiles for the dark matter halo models (Section \ref{darkmatterhalo}). In Section \ref{targetgalaxy}, we describe our target LSBGs, including the selection criteria and stellar mass estimation. We then model the neutrino energy spectrum and calculate the total number of past CCSNe in Section \ref{neutrinoemission}. Finally, we provide the procedure for deriving the constraints on the interaction cross section in Section \ref{derivationcrosssection}.

\subsection{Dark Matter Halo Profiles\label{darkmatterhalo}}
In this study, we consider a scenario in which dark matter density distribution of galaxies transitions from a centrally cusped profile to a flat-cored profile due to energy injection. We adopt the Navarro-Frenk-White (NFW) profile \cite{Navarro:1995iw} for the cusped density profile, and a modified NFW profile with a core (hereafter, the cNFW profile) for the cored density profile, respectively:
\begin{align}
    \rho_{\text{NFW}}(r)&=\frac{\rho_{\text{NFW},0}}{\left(\frac{r}{r_s}\right)\left(1+\frac{r}{r_s}\right)^2}, \\
    \rho_{\text{cNFW}}(r)&=\frac{\rho_{\text{cNFW},0}}{\left(\frac{r_c}{r_s}+\frac{r}{r_s}\right)\left(1+\frac{r}{r_s}\right)^2},
\end{align}
where $r$ is the radius from the galactic center, $\rho_{*,0}$ is the characteristic halo density, $r_s$ is the scale radius, and $r_c$ is the core radius.

The halo mass profile $M(r)$ within a sphere of radius $r$ for the cNFW profile is analytically obtained as follows:
\begin{align}
    \frac{M(r)}{M_0}=\begin{cases}
        \ln(1+\tilde{r})-\frac{\tilde{r}(2+3\tilde{r})}{2(1+\tilde{r})^2} & x=1,\\
        \frac{x^2\ln(1+\tilde{r}/x)+(1-2x)\ln(1+\tilde{r})}{(1-x)^2}-\frac{\tilde{r}}{(1+\tilde{r})(1-x)} & x\neq1,
    \end{cases}
\end{align}
where $M_0=4\pi \rho_0 r^3_s, \tilde{r}=r/r_s, $ and $x=r_c/r_s$. Note that the mass profile for the NFW profile is obtained by taking the limit of $r_c \rightarrow 0$.

We define the virial radius $r_{200}$ as the radius within which the average density is 200 times the critical density of the universe, $\rho_{\text{crit}}=3H^2_0/(8\pi G_N)$. Here, $H_0 =71\, \rm km/s/Mpc$ is the present-day Hubble constant and $G_N$ is Newton's gravitational constant. The virial velocity $V_{200}$ and virial mass $M_{200}$ are then given by:
\begin{align}
    V^2_{200}&=\frac{G_N M_{200}}{r_{200}},\\
    M_{200}&=\frac{800\pi}{3}r^3_{200}\rho_{\text{crit}}.
\end{align}
The characterisitc density parameter of the NFW profile can be obtained using the concentration parameter $c=r_{200}/r_s$,
\begin{align}
    \rho_{\text{NFW},0} = \frac{200}{3}\rho_{\text{crit}}\frac{c^3}{\ln(1+c)-\frac{c}{1+c}}.
\end{align}
The specific values of $c, V_{200}, r_{200}$ and $M_{200}$ for each galaxy are listed in Table \ref{tab:galaxy_properties} (see next section for details). 

\subsection{\label{targetgalaxy}Target Galaxy}
As the targets of our analysis, we utilize LSBGs. LSBGs are very diffuse galaxies with low stellar densities and have a high concentration of dark matter extending into their central regions, which makes them ideal targets for our study to analyze the nature of dark matter.
We selected our galaxy sample from three observational datasets: de Blok et al.~\cite{deBlok:2001rgg}, DensePak integral field spectroscopy \cite{KuziodeNaray:2007qi,KuziodeNaray:2006wh}, and the SPARC catalog \cite{Li:2020iib}. In these datasets, the observational data were analyzed using a pseudo-isothermal (pISO) profile as the cored dark matter model:
\begin{align}
    \rho_{\text{iso}}(r)=\frac{\rho_{\text{iso},0}}{1+(1+r/r_{\text{iso},c})^2},
\end{align}
where $\rho_{\text{iso,0}}$ is the central density and $r_{\text{iso},c}$ is the core radius. Since our analysis assumes the cNFW profile, we performed a profile matching to calibrate the core radius. We optimized the cNFW profile to best match the pISO profile within the radius from the center to $r_{\text{iso},c}$, and thereby modified the core radius $r_{\text{mod},c}$ from the cNFW profile. We define the modification ratio of the core radius as $\kappa = r_{\text{mod},c}/r_{\text{iso},c}$. While the $\kappa$ value for the entire sample of 175 galaxies in the SPARC catalog range from a minimum of $\sim 10^{-2}$ to a maximum of $\sim 10^2$, 76\% of the galaxies fall within a moderate range of $0.1 < \kappa < 1$, as well as all the galaxies from de Blok et al. and DensePak samples.  As described below, we omit galaxies with extreme $\kappa$s. The original core radius $r_{\text{iso},c}$ and the modification ratio $\kappa$ for our selected galaxies are given in column 8 and 9 in Table \ref{tab:galaxy_properties}.

\begin{table*}[ht]
\caption{DM properties for each LSBG. The parameters $c$, $V_\text{200}$, $r_\text{200}$, $M_\text{200}$ are derived from the NFW profile fitting, while $r_c$ and $\rho_{\text{iso},0}$ are obtained from the pseudo-isothermal profile fitting. The first 10 galaxies are taken from de Blok et al.~\cite{deBlok:2001rgg}, the following 4 from DensePak \cite{KuziodeNaray:2007qi,KuziodeNaray:2006wh}, and the final 15 from the SPARC catalog \cite{Li:2020iib}.}
\label{tab:galaxy_properties}
\begin{ruledtabular}
\begin{tabular}{cccccccc}

Name & $c$ & $V_{200}[\mathrm{km/s}]$ & $r_{200}[\mathrm{kpc}]$ & $M_{200}[\mathrm{M}_\odot]$ & $\rho_{\rm iso,\mathrm{0}} [10^{-3}\mathrm{M}_\odot \mathrm{pc}^{-3}]$ & $r_c \pm 1\sigma [\mathrm{kpc}]$ & $\kappa$ \\
\colrule
F563-1 & 9.9 & 88.8 & 125.07 & $2.29 \times 10^{11}$ & $79.0 $ & $1.72 \pm 0.26$ & 0.31\\

F568-3 & 2.3 & 218.6 & 307.89 & $3.42 \times 10^{12}$ & $25.7 $ & $3.07 \pm 0.63$ & 0.17 \\

F579-V1 & 23.2 & 67.9 & 95.63 & $1.03 \times 10^{11}$ & $694.4$ & $0.55 \pm 0.04$ &0.30 \\

F583-1 & 5.4 & 90.6 & 127.61 & $2.44 \times 10^{11}$ & $31.5$ & $2.26 \pm 0.11$ & 0.24 \\

F583-4 & 5.9 & 76.0 & 107.04 & $1.44 \times 10^{11}$ & $80.6$ & $1.02 \pm 0.12$ & 0.17 \\

UGC5750 & 1.9 & 116.9 & 164.65 & $5.23 \times 10^{11}$ & $7.1$ & $4.67 \pm 0.74$ & 0.19 \\

F563-V2 & 6.5 & 136.0 & 191.55 & $8.24 \times 10^{11}$ & $96.6 $ & $1.70 \pm 0.24$ & 0.18 \\

F568-1 & 6.7 & 163.8 & 230.70 & $1.44 \times 10^{12}$ & $90.7 $ & $2.11 \pm 0.15$ & 0.20\\

F568-V1 & 14.6 & 88.8 & 125.07 & $2.29 \times 10^{11}$ & $146.1$ & $1.41 \pm 0.15$ & 0.39\\

F574-1 & 8.2 & 87.1 & 122.68 & $2.17 \times 10^{11}$ & $63.1 $ & $1.74 \pm 0.09$ &0.28\\[1.5ex]

F563-V2 & 7.8 & 104 & 146.48 & $3.69 \times 10^{11}$ & $102 $ & $1.4 \pm 0.2$ & 0.22 \\

F563-1 & 7.0 & 102 & 143.66 & $3.48 \times 10^{11}$ & $54$ & $2.1 \pm 0.2$ & 0.24\\

F583-4 & 5.8 & 86 & 121.13 & $2.08 \times 10^{11}$ & $63$ & $1.2 \pm 0.2$ &0.24 \\

F583-1 & 4.9 & 110 & 154.93 & $4.36 \times 10^{11}$ & $27$ & $2.5 \pm 0.2$ &0.31 \\[1.5ex]

UGCA281 & 4.19  & 87.6 & 123.44 & $2.21\times 10^{11}$ & 173.8 & $0.34 \pm 0.09$ & 0.19\\

UGC04483 & 8.98 & 24.6 & 34.61  & $4.86\times10^{9}$ & 120.2 & $0.32\pm 0.11$ &0.24 \\

NGC6789 & 4.01 & 437.4 & 616.13 & $2.74\times10^{13}$ & 524.8  & $0.70\pm 0.80$ & 0.12  \\

UGC06628 & 19.95 & 28.5 & 40.11 & $7.57\times10^9$  & 426.6 & $0.29\pm0.72$ & 0.31  \\

ESO444-G084 & 7.15 & 78.8 & 111.00  & $1.60\times 10^{11}$ & 173.8 &$0.76\pm 0.36$  & 0.12 \\

UGC05764 & 16.99 & 36.8 & 51.89 & $1.64\times 10^{10}$ & 338.8 & $0.46\pm 0.18$ & 0.21  \\

UGC07866 &  4.56& 44.9 & 63.27  & $2.97\times 10^{10}$ & 56.2 & $0.58\pm 2.59$  & 0.20 \\

F579-V1 & 21.76 & 64.9 & 91.41 & $8.96\times 10^{10}$ & 691.8 & $0.59\pm 0.67$ & 0.21\\

UGCA444 & 1.01 & 219.5 & 309.21 & $3.47\times 10^{12}$ & 55.0 & $0.69\pm 0.13$ & 0.19  \\

UGC04325 & 17.39  & 58.3 & 82.17 & $6.51\times 10^{10}$ &363.1  & $0.67\pm 0.44$ & 0.23 \\

UGC07261 & 15.96 & 51.5 & 72.55 & $4.48\times 10^{10}$  & 295.1  & $0.59\pm 0.91$ & 0.27 \\

F583-4 & 3.83 & 91.8 & 129.34 & $2.54\times 10^{11}$ & 53.7 & $1.22\pm 0.80$ & 0.14 \\

F563-V2 & 6.08 & 135.7 & 191.17 &  $8.19\times 10^{11}$& 120.2 & $1.68\pm 2.21$ & 0.10 \\

UGC08550 & 11.89 & 43.7 & 61.54 & $2.73\times 10^{10}$ & 147.9 & $0.64\pm 0.49$ & 0.27  \\

D512-2 & 5.86  & 42.7  & 60.08 &  $2.54\times 10^{10}$& 45.7 & $0.84\pm 2.62$  & 0.23 \\

\end{tabular}
\end{ruledtabular}
\end{table*}

In all three datasets, several galaxies feature a best-fit concentration parameter fixed at $c=1.0$. These correspond to systems where the baryon component accounts for the majority of the observed rotation velocity, causing the NFW profile fit to fail and yield unphysically small or negative values for $c$, which were artificially constrained to unity. Such galaxies were omitted from our analysis to maintain the robustness of our dark matter profile modeling.

The specific selection criteria and stellar mass estimation methods for each dataset are as follows:
\begin{itemize}
    \item de Blok et al.~sample: we selected galaxies from SAMPLE I, which has a full set of photometry and HI data, and adopted the model with a constant stellar mass-to-light ratio of $\Upsilon_\star(R)=1.4$. We excluded F571-8 and UGC6614 due to their significant bulge contributions, leaving a total of 10 galaxies. The total stellar mass $M_\star$ was estimated using the relation $M_\star = \Upsilon_\star(R)\times 10^{-0.4(M_{\rm abs}(R)-M_{\rm abs,\odot}(R))}$, adopting the solar R-band absolute magnitude $M_{\rm abs, \odot}(R)=4.4$. R-band absolute magnitude ($M_{\rm abs}(R)$) was converted from the B-band magnitude assuming a typical luminosity-weighted color index of $B-R=0.9$ \cite{de1995surface}.

    \item DensePak sample: we utilized the {\tt popsynth} model and selected 4 galaxies that have complete concentration and multicolor photometry data. The stellar mass was derived using the exponential disk integration formula $m = \mu_0 - 2.5 \log_{10}(2\pi \alpha^2)$, where $m$ is the total apparent magnitude, $\mu_0$ is the central surface brightness and $\alpha$ is the angular scale length. Then $M_\star$ was computed using the individual mass-to-light ratio $\Upsilon_{\star}$ and the absolute R-band magnitude converted from the apparent magnitude.
    
    \item SPARC sample: for the initial cuspy halo, we adopted the best-fit parameters of the NFW profile derived under flat priors. Out of 175 galaxies, we selected our LSBG sample by requiring B-band surface birghtness to be $\mu_B \ge 23.0 \, \rm mag/arcsec^2$ \cite{bothun1997low}. This value was converted from the effective surface brightness at 3.6 $\mu \rm{m}$ assuming a color index of $B-[3.6]=2.7$ \cite{Schombert:2013hga} and solar absolute magnitude of 3.24 at $3.6\mu \rm m$ \cite{Oh:2008ww}, as adopted in the SPARC mass models \cite{Lelli:2016zqa}. Furthermore, we excluded galaxies with an extreme core modification ratio $\kappa < 0.1$ and $\kappa > 1$. From the remaining galaxies, we selected the top 15 with the smallest core radii ($r_{\text{mod},c}$). 
\end{itemize}

\begin{table*}[ht]
\caption{Galaxy paramters calculated for each LSBG. The columns represent the total stellar mass $M_\star$, the estimated number of core-collapse CCSNe $N_{\rm CCSNe}$ using Eq.(\ref{CCSNe}), the total energy released by these CCSNe $E_{\rm tot}$, the required energy injected into the dark matter halo to form the observed core $E_{\rm inj}$, and the fraction of the total neutrino energy injected to dark matter halo $\epsilon = E_{\rm inj}/E_{\rm tot}$.}
\label{tab:energetic_properties}
\begin{ruledtabular}
\begin{tabular}{cccccc}

Name & $M_{\mathrm{\star}}[\mathrm{M_\odot}]$ & $N_{\mathrm{CCSNe}}$ & $E_{\mathrm{tot}}[\mathrm{MeV}]$ & $E_{\mathrm{inj}}[\mathrm{MeV}]$ & $\mathrm{\epsilon}$ \\
\colrule
F563-1 & $1.54\times10^9$ & $1.71\times 10^7$ & $3.20\times10^{66}$ & $3.94\times 10^{62} $ & $1.23\times 10^{-4}$\\

F568-3 & $3.86\times 10^9$ & $4.29\times 10^7$ & $8.04\times 10^{66}$ & $3.35\times 10^{63} $ & $4.16\times 10^{-4}$\\

F579-V1 & $6.11\times10^{9}$ & $6.80\times10^7$ & $1.27\times10^{67}$ & $1.36\times 10^{62} $ & $1.07\times 10^{-5}$\\

F583-1 & $7.35\times10^{8}$ & $8.18\times10^6$ & $1.53\times10^{66}$ & $2.00\times 10^{62} $ & $1.31\times 10^{-4}$\\

F583-4 & $1.06\times10^{9}$ & $1.18\times10^7$ & $2.21\times10^{66}$ & $4.11\times 10^{61} $ & $1.86\times 10^{-5}$\\

UGC5750 & $5.57\times10^{9}$ & $6.20\times10^7$ & $1.16\times10^{67}$ & $3.36\times 10^{62} $ & $2.89\times 10^{-5}$\\

F563-V2 & $3.52\times10^{9}$ & $3.91\times10^7$ & $7.33\times10^{66}$ & $8.34\times 10^{62} $ & $1.14\times 10^{-4}$\\

F568-1 & $3.21\times10^{9}$ & $3.57\times10^7$ & $6.69\times10^{66}$ & $2.40\times 10^{63} $ & $3.59\times 10^{-4}$\\

F568-V1 & $2.67\times10^{9}$ & $2.97\times10^7$ & $5.56 \times 10^{66}$ & $6.33 \times 10^{62} $ & $1.14 \times 10^{-4}$\\

F574-1 & $4.23\times 10^{9}$ & $4.71\times 10^7$ & $8.81 \times 10^{66}$ & $2.48\times 10^{62} $ & $2.82\times 10^{-5}$\\[1.5ex]

F563-V2 & $1.98\times 10^{9}$ & $2.20\times 10^7$ & $4.12\times 10^{66}$ & $3.50\times 10^{62} $ & $8.50\times 10^{-5}$\\

F563-1 & $1.50 \times 10^{9}$ & $1.67\times 10^7$ & $3.12\times 10^{66}$ & $4.23\times 10^{62} $ & $1.36\times 10^{-4}$\\

F583-4 & $8.22 \times 10^{8}$ & $9.15 \times 10^6$ & $1.71\times 10^{66}$ & $1.09\times 10^{62} $ & $6.39\times 10^{-5}$\\

F583-1 & $2.56 \times 10^{8}$ & $2.85\times 10^6$ & $5.34\times 10^{65}$ & $5.48\times 10^{62} $ & $1.03\times 10^{-3}$\\[1.5ex]

UGCA281 & $9.51\times 10^7$ & $1.06\times 10^6$ & $1.98\times 10^{65}$ & $2.17\times 10^{61}$ & $1.09\times 10^{-4}$ \\

UGC04483 & $6.50\times 10^6$ & $7.24\times 10^{4}$ & $1.35\times 10^{64}$ & $3.78\times 10^{59}$ & $2.79\times 10^{-5}$ \\

NGC6789 & $5.00\times 10^7$ & $5.57\times 10^5$ & $1.04\times 10^{65}$ & $2.74\times 10^{64}$ &  $2.63\times 10^{-1}$ \\

UGC06628 & $1.87\times 10^9$ & $2.08\times 10^7$ & $3.90\times 10^{66}$ &$4.49\times 10^{60}$ & $1.15\times 10^{-6}$  \\

ESO444-G084 & $3.62\times 10^7$ & $4.03\times 10^5$  & $7.55\times 10^{64}$ & $4.34\times 10^{61}$ &$5.75\times 10^{-4}$ \\

UGC05764 &  $4.16\times 10^7$& $4.64\times 10^5$ & $8.68\times 10^{64}$ &$5.58\times 10^{60}$& $6.43\times 10^{-5}$   \\

UGC07866 & $6.20\times 10^7$ & $6.90\times 10^5$  &$1.29\times 10^{65}$  & $9.89\times 10^{60}$ &  $7.65\times 10^{-5}$\\

F579-V1 & $6.04\times 10^{9}$ & $6.73\times 10^{7}$ & $1.26\times 10^{67}$ & $1.42\times 10^{62}$ &  $1.13\times 10^{-5}$ \\

UGCA444 &$6.00\times 10^6$  &$6.68\times 10^4$  & $1.25\times 10^{64}$ & $4.27\times 10^{62}$ &$3.42\times 10^{-2}$ \\

UGC04325 & $1.05\times 10^9$ & $1.17\times 10^7$ & $2.20\times 10^{66}$ & $6.71\times 10^{61}$ & $3.05\times 10^{-5}$\\

UGC07261 & $8.24\times 10^8$ & $9.17\times 10^6$ & $1.72\times 10^{66}$ & $5.12\times 10^{61}$ &$2.98\times 10^{-5}$ \\

F583-4 & $8.75\times 10^8$ & $9.74\times 10^6$ & $1.82\times 10^{66}$ &  $7.56\times 10^{61}$& $4.15\times 10^{-5}$ \\

F563-V2 & $1.52\times 10^{9}$ & $1.70\times 10^7$ & $3.17\times 10^{66}$ & $8.66\times 10^{62}$ & $2.73\times 10^{-4}$ \\

UGC08550 & $1.44\times 10^8$ &$1.61\times 10^6$  & $3.01\times 10^{65}$ & $1.44\times 10^{61}$ &  $4.78\times 10^{-5}$\\

D512-2 & $1.62\times 10^8$ & $1.81\times 10^6$ & $3.39\times 10^{65}$ &  $1.23\times 10^{61}$& $3.64\times 10^{-5}$\\

\end{tabular}
\end{ruledtabular}
\end{table*}

\subsection{\label{neutrinoemission}Neutrino Emission from Core-Collapse Supernovae}
We consider neutrinos emitted from CCSNe within the target galaxies. In a CCSN, the gravitational binding energy is released across all neutrino flavors. Detailed numerical simulations indicate that the average energies differ by $\sim10\%$ among different species, and their luminosities can vary by up to a factor of two \cite{Keil:2002in}. However, for simplicity, we assume each flavor equally shares the total energy and the spectra are identical, as was done in previous studies \cite{Heston:2024ljf}. We adopt a time-integrated pinched Fermi-Dirac distribution for the total neutrino energy spectrum $F_\nu(E_\nu)$ \cite{Keil:2002in}:
\begin{align}
    F_\nu(E_\nu)= \frac{E_{\nu, \rm SN}}{\langle E_\nu \rangle^2}\frac{(\alpha+1)^{\alpha+1}}{\Gamma(\alpha+1)}\left(\frac{E_\nu}{\langle E_\nu \rangle}\right)^\alpha \exp\left(-\frac{(\alpha+1)E_\nu}{\langle E_\nu \rangle} \right),
\end{align}
where $E_{\nu,\rm SN}$ is the total energy released in neutrinos, $\langle E_\nu \rangle$ is the average neutrino energy, and $\alpha$ is the pinching parameter. This is valid since the typical duration of neutrino emissiono ($\sim 10$s) is negligibly short compared to the timescale of dark matter core formation. We adopt $E_{\nu,\rm SN}=3\times 10^{53}$erg, $\langle E_\nu \rangle =15$MeV, and $\alpha = 2.3$ \cite{Heston:2024ljf}.

We estimate the total number of historical CCSNe ($N_{\rm CCSNe}$) based on the current stellar mass $M_\star$ of each galaxy. Assuming a stellar initial mass function (IMF) $\xi(m) \propto m^{-\gamma}$, $N_{\rm CCSNe}$ is obtained by integrating the IMF over the mass range of the progenitor stars:
\begin{align}
    N_{\rm CCSNe} &= M_\star \frac{\int_{8M_\odot}^{100M_\odot}\xi(m)dm}{\int_{0.1M_\odot}^{100M_\odot}m \xi(m)dm}.\label{CCSNe}
\end{align}
In this study, we adopt the Kroupa IMF \cite{Kroupa:2002ky}, where the index $\gamma$ is given by $\gamma = 0.3$ for $m\leq 0.08M_\odot$, $\gamma=1.3$ for $0.08M_\odot < m \leq 0.5M_\odot$, and $\gamma=2.3$ for $0.5M_\odot < m$. Here, we assume that stars with initial masses between $8M_\odot$ and $100M_\odot$ undergo core-collapse. Finally, the total neutrino energy is calculated as $E_{\rm tot}=N_{\rm CCSNe}\times E_{\nu,\rm SN}$. The derived values of $M_\star$, $N_{\rm CCSNe}$, and $E_{\rm tot}$ for each galaxy are summarized in Table \ref{tab:energetic_properties}.

\begin{figure}[ht]
    \centering
    \includegraphics[width=\linewidth]{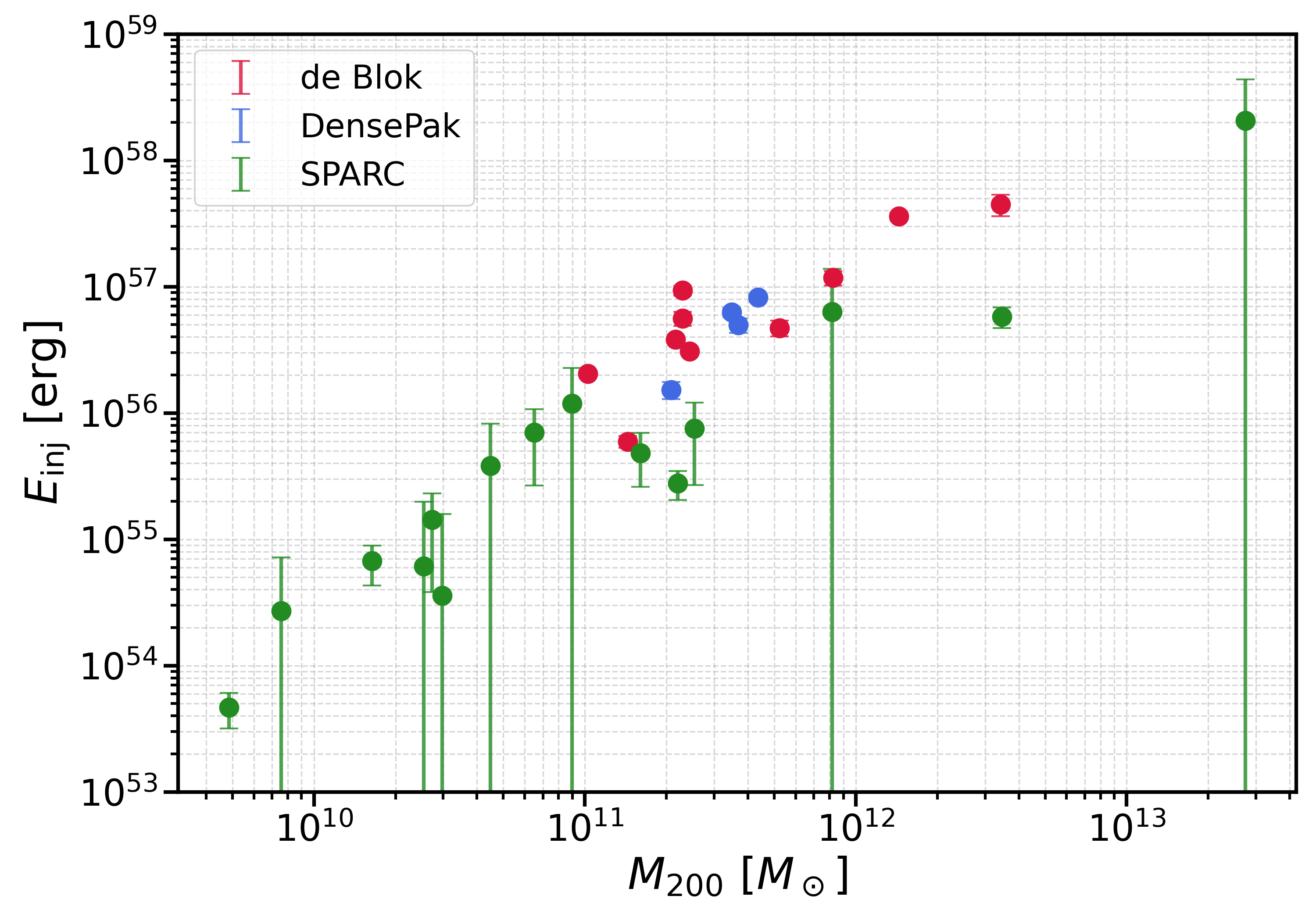}
    \caption{The energy required for dark matter core formation ($E_{\rm int}$) as a function of the halo virial mass ($M_{\rm 200}$). The data points from the de Blok, DensePak, and SPARC datasets are shown in red, blue, and green, respectively. The circles represent the energy evaluated at the observed core radius $r_c$, while the upper and the lower error bars correspond to the values evaluated at $r_c+1\sigma$ and $r_c-1\sigma$. For galaxies where the lower bound of the core radius is negative, the lower error bars are truncated at zero.}
    \label{DeltaW}
\end{figure}

\subsection{\label{derivationcrosssection}Derivation of Constraints on the Cross Section}
To derive the constraints on the $\nu$-DM scattering cross section, we follow the previous work calculated for dwarf spheroidals by Heston et al.~(2024) \cite{Heston:2024ljf}, with a modification to the counting of scatterings: we evaluate the number of DM particles that scatter at least once, rather than the total number of scattering events, so that each DM particle is counted at most once (see \S \ref{massloss}). We consider the scenario where the energy injected by CCSN neutrinos into the dark matter halo drives the formation of the observed cores. The derivation is divided into two regimes, separated by a threshold dark matter mass, $m_{\chi, \rm lim}$.

In the low mass regime (\S \ref{massloss}), the dark matter particles can acquire enough kinetic energy from a single neutrino scattering to escape from the galactic gravitational potential. In this regime, we consider that core formation occurs due to the reduction of dark matter particles, and we derive the cross section constraints by evaluating this mass loss process. Conversely, in the regime heavier than $m_{\chi, \rm lim}$, the total mass of the halo is conserved during the core formation process and we derive the cross section by evaluating the difference in the gravitational potential energy of the halo between the cuspy and cored profiles (\S \ref{energyloss}).

This threshold mass is determined by the condition where the spectrum-averaged maximum kinetic energy transferred to a dark matter particle exceeds the kinetic energy required to escape the galactic gravitational potential:
\begin{align}
    \frac{1}{2}m_\chi v^2_{\rm esc} \leq \langle \Delta E_{\rm int}\rangle.
\end{align}
Assuming the dark matter particle is initially at rest, the maximum energy transferred in a single relativistic collision is given by $ \Delta E_{\rm int}=2E^2_\nu/(2E_\nu + m_\chi)$. Here, $v_{\rm esc}$ is the escape velocity from the center of the galaxy, and $\langle \dots \rangle$ denotes the average over the CCSN neutrino spectrum. For our LSBG sample, the escapse velocities range from approximately $\sim 90$ to $1400$ km/s. Compared to the previous study on dSphs \cite{Heston:2024ljf}, these values are higher, which consequently lead to lower mass thresholds in the range of $7$ to $117$ GeV.

\subsubsection{Mass loss regime ($m_\chi \leq m_{\chi, \rm lim}$)}\label{massloss}
In the low mass regime, we derive the cross section bound by estimating the maximum mass deficiency $\Delta M_{\rm lim}$ required to form the observed core:
\begin{align}
    \Delta M_{\rm lim} = M_{\rm NFW}(r_c) - M_{\rm cNFW}(r_c).
\end{align}

We assume that all CCSNe occur at the galactic center and that neutrinos are emitted isotropically. The mean number of scatterings that a DM particle experience at radius $r$ is 
\begin{align}
    \lambda(r) = \frac{\sigma_{\nu\chi}N_{\nu,\rm tot}}{4\pi r^2}.
\end{align}
Here, $N_{\nu, \rm tot}$ is given by the multiplication of $N_{\rm CCSNe}$ and $N_\nu$, where $N_\nu$ is the total number of neutrinos emitted per single CCSN event, calculated as $N_\nu = \int_{0\rm GeV}^{1 \rm GeV} F_\nu dE_\nu$. Since CCSNe neutrinos scatter off a DM particle independently with a small probability, the number of scatterings $X$ follows a Poisson distribution,
\begin{align}
    P(X=k) = \frac{\lambda^k e^{-\lambda}}{k!}.
\end{align}

In previous work \cite{Heston:2024ljf}, the number of scattered particles was estimated as the total number of scattering events, $\sigma_{\nu\chi}N_{\nu,\rm tot}\int \rho_{\rm NFW} dr$. This is valid only when the average number of scatterings is below one $\lambda(r) \ll1$. Because this condition is violated near the galactic center, we instead count the DM particles that scatter at least once, with probability $P(X\ge1)=1-e^{-\lambda(r)}$. The mass removed from the core is
\begin{align}
    M_{\rm removed}(\sigma_{\nu\chi}) = \int_{0}^{r_c}4\pi r^2 \rho_{\rm NFW}(r)[1-e^{-\lambda(r)}]dr,
\end{align}
which reduces to the expressions in \cite{Heston:2024ljf} for $\lambda \ll 1$. The constraint on the cross section in this regime is obtained as the value of $\sigma_{\nu\chi}$ for which the removed mass coincides with $\Delta M_{\rm lim}$:
\begin{align}
    M_{\rm removed }(\sigma_{\nu\chi})=\Delta M_{\rm lim}.
\end{align}
Since both $\Delta M_{\rm lim}$ and $M_{\rm removed}(\sigma_{\nu\chi})$ are independent of $m_\chi$, $\sigma_{\nu\chi}$ is constant for $m_\chi < m_{\chi, \rm lim}.$

\begin{figure*}[t] 
    \centering
    \begin{minipage}{0.48\textwidth}
        \centering
        \includegraphics[width=\linewidth]{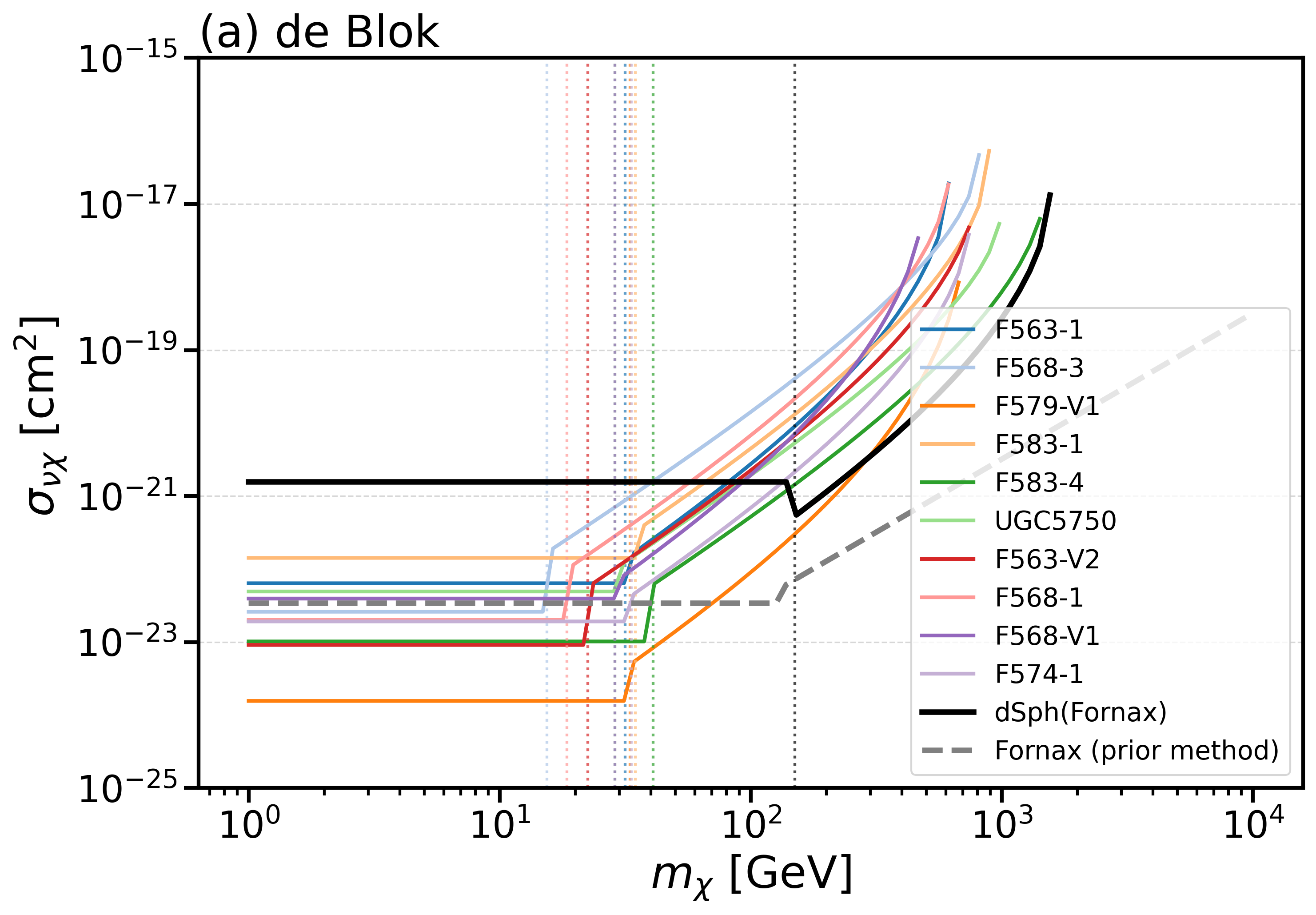}
    \end{minipage}\hfill
    \begin{minipage}{0.48\textwidth}
        \centering
        \includegraphics[width=\linewidth]{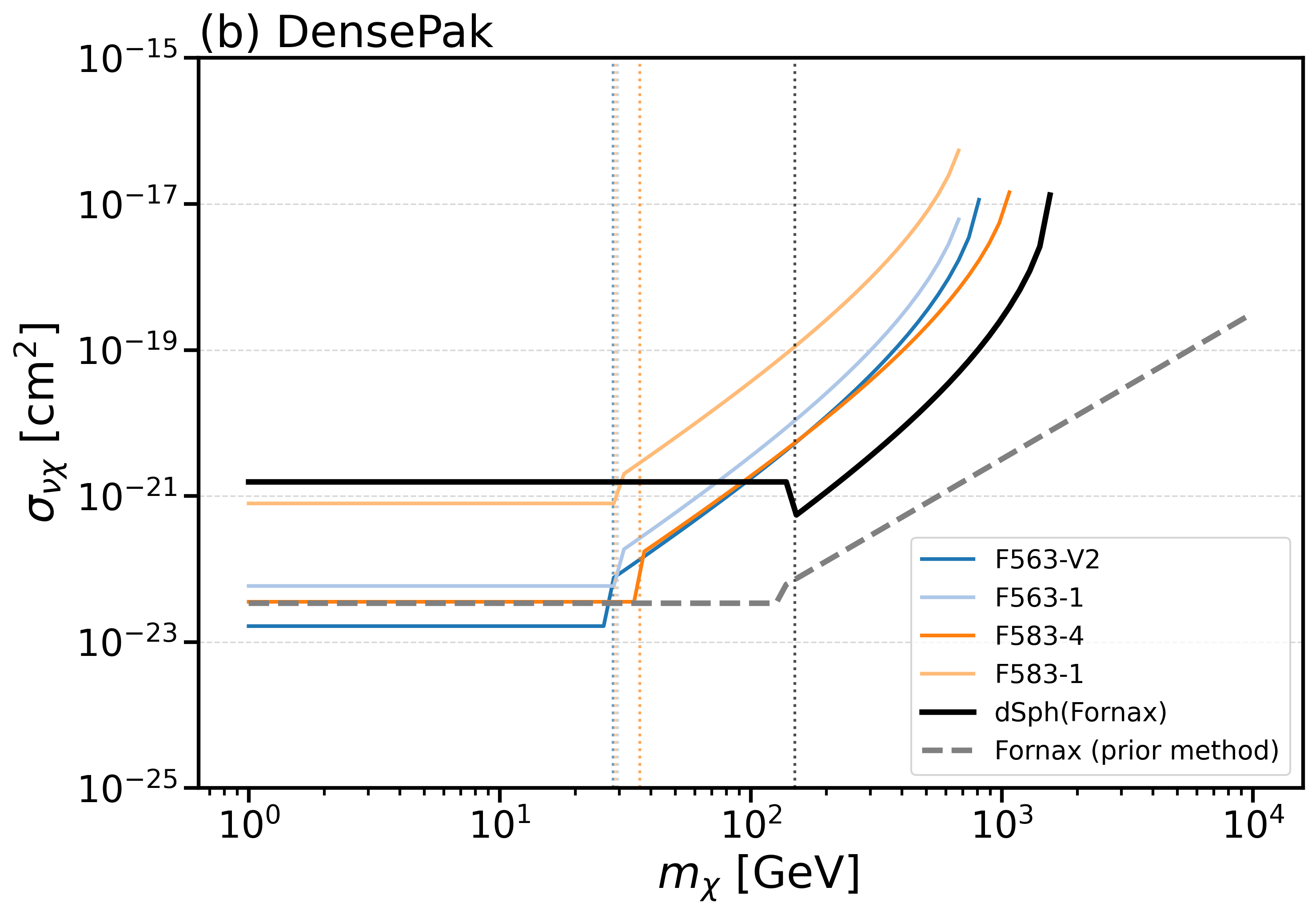}
    \end{minipage}

    \begin{minipage}{0.48\textwidth}
        \centering
        \includegraphics[width=\linewidth]{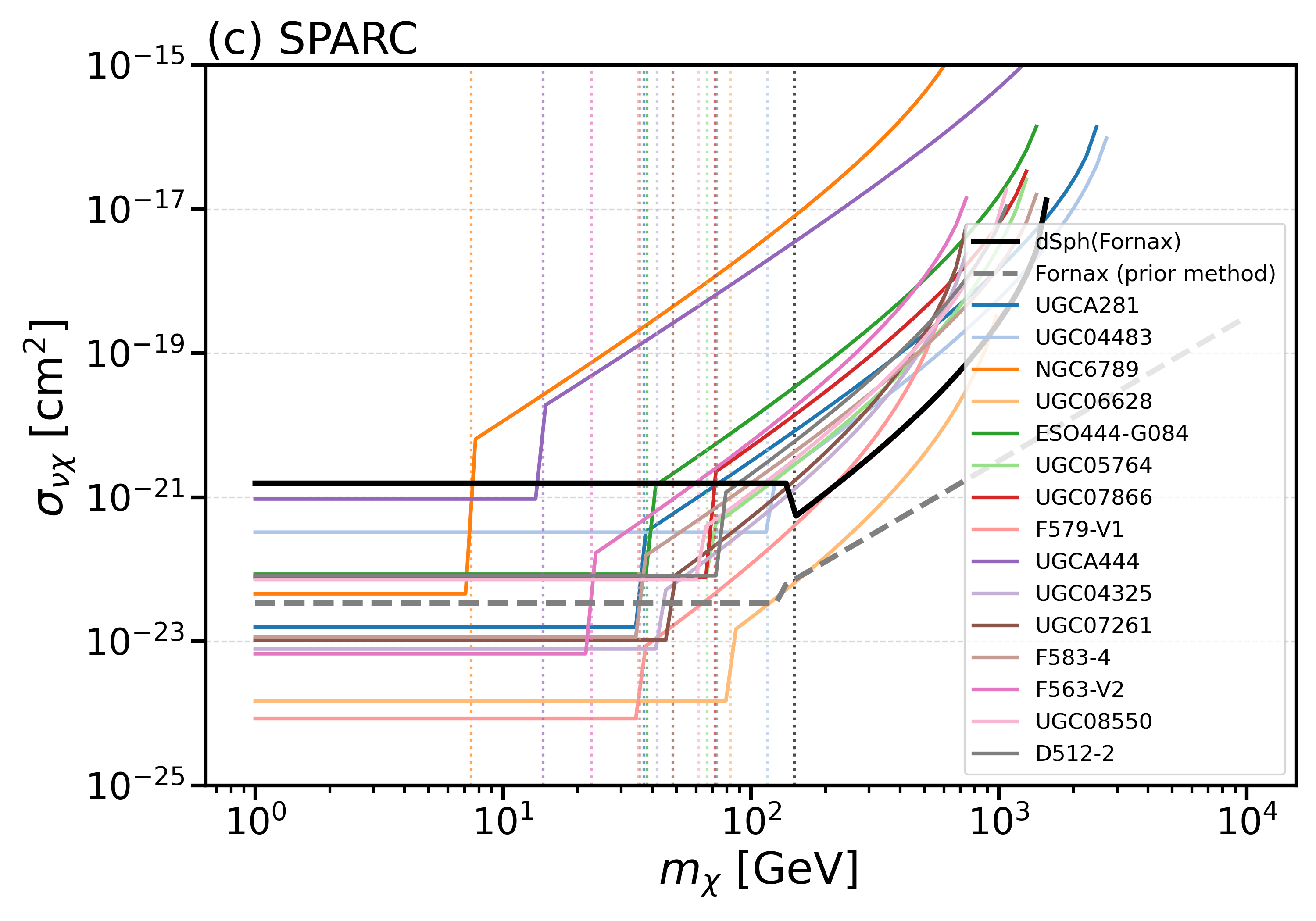}
    \end{minipage}\hfill
    \begin{minipage}{0.48\textwidth}
        \vspace{0pt}
        \small
        \setlength{\baselineskip}{1.2em}
        \caption{Upper limits on the dark matter-neutrino scattering cross section as a function of the dark matter mass. The three panels display the results derived from the different LSBG datasets: (a) de Blok, (b) DensePak, and (c) SPARC. Each colored line represents the constraint obtained from an individual galaxy within the respective dataset. The black solid line shows the constraint from the Fornax dSph for comparison. The gray dashed line shows the constraint from Fornax dSph from \cite{Heston:2024ljf}. The vertical dotted lines indicate $m_{\chi,\rm lim} $ for each galaxy.}
        \label{fig1} 
    \end{minipage}
\end{figure*}

\subsubsection{Energy injection regime ($m_\chi > m_{\chi, \rm lim}$)}\label{energyloss}
In the high mass regime, the dark matter particles do not acquire enough energy to escape from the galaxy. Instead, they remain gravitationally bound and form cores while conserving the virial mass $M_{\rm 200}$. In this scenario, we derive the upper bound on the cross section by equating the energy injected by the scatterings at certain cross section with the energy needed to form the observed core. From the virial theorem, the latter energy is given by $\Delta W/2$ \cite{Penarrubia:2012bb}, where $\Delta W$ is the difference in the gravitational potential energy between the respective profiles,
\begin{align}
  \Delta W = W_{\rm cNFW} - W_{\rm NFW},
\end{align}
where
\begin{align}
  W = -4\pi G\int_0^{r_{200}} r\rho(r)M(r)dr.
\end{align}
We evaluate $W_{\rm cNFW}$ using the upper bound of the core radius $r_{\rm upper}=r_c+1\sigma,$ which makes the cross section limits conservative, since a larger core requires a larger energy injection $\Delta W$ and hence a larger cross section. 
As in the low-mass regime, we count each DM particle at most once. In this regime, however, this is a conservative assumption for simplicity, rather than a physical restriction, since a bound DM particle can receive additional energy from subsequent scatterings. The DM mass within $r_{\rm 200}$ that undergoes at least one scattering is written as 
\begin{align}
    M_{\rm scattered}(\sigma_{\nu\chi})=\int_{0}^{r_{200}}4\pi r^2 \rho_{\rm NFW}(r)[1-e^{-\lambda(r)}]dr.\label{scattered}
\end{align}
Assuming that each scattered DM particle receives the energy of a single scattering, $\langle \Delta E_{\rm int}\rangle$, the total energy injected into the halo is $\langle \Delta E_{\rm int}\rangle M_{\rm scattered}/m_{\chi}$. By requiring that this energy corresponds to $\Delta W/2$, the constraint in this regime is obtained as the value of $\sigma_{\nu\chi}$ that satisfies
\begin{align}
    M_{\rm scattered}(\sigma_{\nu\chi}) = \frac{\Delta W/2}{\langle \Delta E_{\rm int}\rangle}m_\chi.\label{sigma_2}
\end{align}
Since the energy transferred per scaterring is approximately inversely proportional to the DM mass $\Delta E_{\rm int} \propto 1/m_\chi$, the required scattered mass increases as $M_{\rm scattered} \propto m^2_\chi$. Thus, the constraints in this regime increase with $m_\chi$, as $M_{\rm scattered}$ increases monotonically with $\sigma$.

Since $M_{\rm scattered}$ cannnot exceed the total DM mass within $r_{\rm 200}$, $M_{\rm 200}$, there is a maximum DM mass for which the constraint can be obtained. For the case of one scattering per DM particle, this maximum DM mass $m_{\rm max}^{(1)}$ is given by 
\begin{align}
    m_{\rm max}^{(1)} \sim 2\sqrt{\langle E_\nu^2\rangle}\sqrt{\frac{M_{\rm 200}}{\Delta W}}.
\end{align}
If multiple scatterings per DM particle are allowed, $m_{\rm max}^{(1)}$ increases with the maximum number of scatterings, and the constraint is obtained over a wider range of $m_\chi$ (see \S \ref{sec:multiple}).

\section{Results \& Discussion}\label{results}

Figure \ref{DeltaW} shows the relationship between the virial mass $M_{\rm 200}$ and the required energy injection into the halo to form the core, $E_{\rm inj}$, for each galaxy. The circles and the upper and the lower error bars represent the injected energy calculated using $r_c$ and $r_c \pm 1\sigma$ respectively. Across all three datasets, the required energy generally follows a consistent scaling relation with the halo mass. Furthremore, this result is broadly consistent with the extrapolation to the more massive LSBG regime of the relation derived by Peñarrubia et al.~(2012) \cite{Penarrubia:2012bb}, who calculated the energy required for core formation in the Milky Way's dSphs. We note that for several galaxies in the SPARC dataset, the lower bound of the required energy extends to zero. This corresponds to galaxies where the lower 1$\sigma$ limit of the core radius are negative, which means their dark matter halos are statistically consistent with having no core at all. In such cases, the initial NFW profile remains a valid assumption and no energy injection from neutrino scattering is required to explain the data, which is represented by the error bars extending down to zero.

Figure \ref{fig1} shows the upper limits on the scattering cross section obtained for each of the three datasets. 
The most stringent constraint among all three LSBG samples is provided by the galaxy F579-V1 and UGC06628 from the SPARC dataset. For the low-mass regime ($m_\chi \leq 35\,\rm GeV$), the galaxy F579-V1 provides the strongest upper limit, yielding a constant cross section of 
\begin{align}
    \langle \sigma_{\nu\chi}(m_{\chi}\leq 35\,\rm GeV)\rangle \sim 8.5\times 10^{-25}\rm cm^2.
\end{align}
On the other hand, in the high-mass regime ($m_\chi > 35\,\rm GeV$), the galaxy UGC06628 provides the most stringent constraint. For this galaxy, the threshold mass and the maximum mass are $m_{\chi, \rm lim}=82\,\rm GeV$ and $m^{(1)}_{\rm max} = 1050\,\rm GeV$, yielding the limit of 
\begin{align}
    \langle \sigma_{\nu\chi}(m_{\chi}\leq 82\,\rm GeV)\rangle &\sim 1.5\times 10^{-24}\,\rm cm^2\\
    \langle \sigma_{\nu\chi}(m_\chi=100\,\rm GeV)\rangle &\sim 2.1\times 10^{-23}\,\rm cm^2
\end{align}
for each regime. Compared to F579-V1, UGC06628 has smaller $c$ and $r_{\rm 200}$, resulting in a smaller escape velocity, which extends the constant cross section regime up to a higher $m_\chi$.

Across all datasets, the threshold mass $m_{\chi,\rm lim}$ is typically on the order of several tens of GeV, which is smaller compared to the value of $\sim 100\,\rm GeV$ previously found for dSphs \cite{Heston:2024ljf}. 
For comparison, black solid line and gray dashed line represent the constraints for Fornax dSph, using our method and the original method of \cite{Heston:2024ljf}, respectively. We show this comparison only for Fornax, since it was found to have the most stringent constraint among dSphs explored in \cite{Heston:2024ljf}. Using our method, we find that the constraint for $m_\chi<m_{\chi,\rm lim}=130\,\rm GeV$ becomes $\sigma_{\nu\chi}=1.56\times 10^{-21}\,\rm cm^2$ (black solid line), which is weaker by a factor of $\sim 46$ than that of the previous work, $3.4\times 10^{-23}\,\rm cm^2$ (gray dashed line). This weakening is explained as follows. To form the observed core of $r_c=297\,\rm pc$ for Fornax, $\Delta M_{\rm lim}=8.8 \times 10^6 \,M_\odot$ must be removed, which is 62\% of the DM mass within $r_c$. However, we can only remove up to $7.0\times 10^5\,M_\odot$, assuming the previous value of the cross section and counting each DM particle at most once. In the previous method, this deficit was compensated by effectively counting multiple scatterings of the same DM particles in the inner region. Keeping the assumption that each DM particle scatters at most once, a larger cross section is required, and the constraint becomes $\sim 46$ times larger than the previous value.

Comparing our best LSB constaints to the best dSph constraints (Fornax, black solid), the LSB results represent an improvement by approximately three orders of magnitude in the constant cross section regime. The reason why these two galaxies provide the strongest constraints is a combination of factors: the abundance of neutrino sources, the density of the target dark matter, and the size of the core. First, both F579-V1 and UGC06628 possess large estimated stellar mass of $M_\star =6.04\times 10^{9}M_\odot$ and $1.87\times 10^9M_\odot$, respectively. This means that a massive amount of neutrinos has been produced by past CCSNe, acting as an abundant energy source. Second, their halo concentration parameters are rather high at $c=21.76$ and $19.95$, resulting in a highly dense central dark matter cusp and large dark matter column densities. Finally, the observed core radii are relatively small at $r_c=0.59\,\rm kpc$ and $0.29 \, \rm kpc$, despite the abundant neutrino sources and dense target dark matter, which ultimately leads to these highly stringent limits.

As shown in Figure \ref{fig1}, the constraint is discontinuous at $m_\chi = m_{\chi,\rm lim}$, and the ratio of constraints at this mass is typically larger for LSBGs than for dSphs. For example, the ratio is $\sim 0.3$ for the Fornax dSph, whereas it is $\sim 8$ for F579-V1. Although this gap arises simply from switching between the two different methods at $m_{\chi,\rm lim}$, there is a physical reason why its magnitude varies among galaxies. For $m_\chi>m_{\chi,\rm lim}$, core formation requires an energy injection equivalent to the change in the total gravitational energy of the halo. In contrast, for $m_\chi < m_{\chi, \rm lim}$, scattered particles escape from the galaxy, meaning the core is formed simply by removing the mass difference within $r_c$. Since the difference between the NFW and cNFW profiles extends beyond $r_c$, the energy injection regime accounts for this outer change, whereas the mass loss regime does not. This contribution becomes relatively more significant when the core is smaller compared to the halo. Therefore, the ratio of constraints at $m_{\chi,\rm lim}$ is primarily determined by $\eta=r_c/r_s$, with a smaller $x$ resulting in a larger ratio. Indeed, $\eta=0.23$ for the Fornax dSph, whereas F579-V1 has a significantly smaller value of $\eta=0.03$.

\subsection{Effects of multiple scatterings}\label{sec:multiple}
So far, we have derived the upper limits on the cross section asuuming that each DM particle is scattered by neutrinos at most once. In reality, however, it is possible that a DM particle, after being scattered once, is scattered again by neutrinos from subsequent CCSNe while it is escaping from the galaxy or having moved outward. Here, we examine how the constraints change when multiple scatterings per DM particle are allowed.

In the low-mass regime ($m_\chi < m_{\chi, \rm lim}$), multiple scatterings do not change the constraints, because a DM particle has already obtained enough energy to escape from the galaxy in the first scattering, and subsequent scatterings do not contribute to core formation. On the other hand, in the high-mass regime, each scattering contributes to core formation as long as the DM particle remains in the galaxy. The constraint in this case is obtained by simply replacing $P(X\ge 1 )=1-e^{-\lambda(r)}$ in Eq.~(\ref{scattered}) by the expected value of $X$ with an upper limit of $n$, $E[\text{min} (X,n)]=n-e^{-\lambda}\sum_{k=0}^{n-1}(n-k)\lambda^k/k!$,

\begin{align}
    M_{\rm scattered}(\sigma_{\nu\chi}) = \int_{0}^{r_{200}}4\pi r^2 \rho_{\rm NFW}(r) E[\text{min} (X,n)] dr.
\end{align}
Since the scattered mass can now be as large as $n M_{200}$, a constraint is obtained as long as $M_{\rm scattered}$ does not excced $n M_{200}$, and the maximum DM mass increases as $m^{(n)}_{\rm max}=\sqrt{n}m^{(1)}_{\rm max}$.

\begin{figure}[t]
    \centering
    \includegraphics[width=\linewidth]{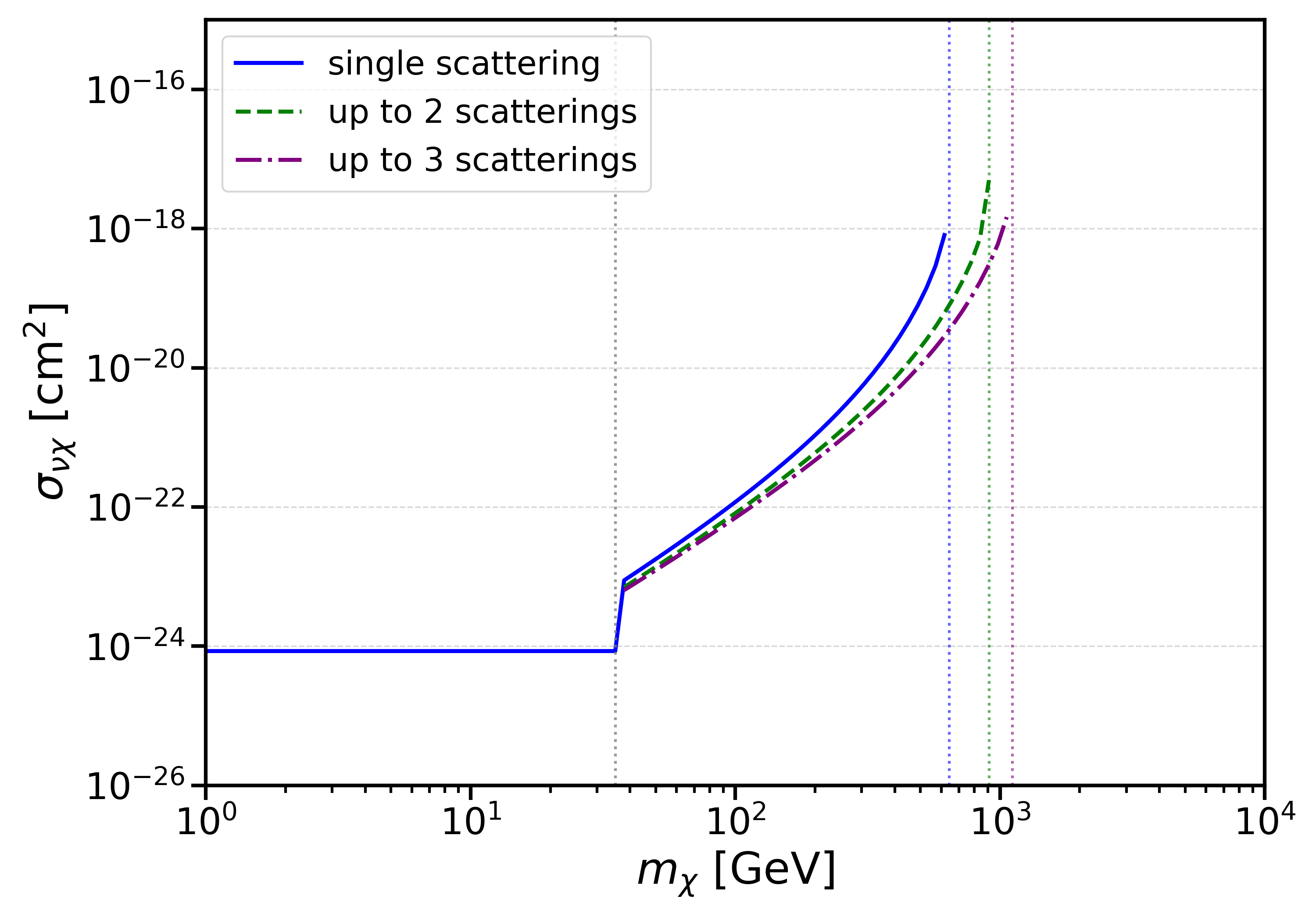}
    \caption{Upper limits on the $\nu-\rm DM$ cross section for F579-V1 when multiple scatterings are allowed. The blue solid, green dashed, and purple dash-dotted lines show the constraints for single scattering, up to two, and up to three scatterings per DM particle, respectively. The dotted vertical lines indicate the maximum DM mass that the constraints can be obtained for each case, $m^{(n)}_{\rm max}=\sqrt{n}m^{(1)}_{\rm max}$.}
    \label{multiple}
\end{figure}

Figure \ref{multiple} shows the effects of multiple scatterings for F579-V1 from the SPARC sample. As the maximum number of scatterings increases, the energy that DM particles can receive from neutrinos increases, and thus the constraint becomes stronger. Since the required $\sigma_{\nu\chi}$ increases with $m_\chi$, more DM particles are scattered more than once at larger $m_\chi$, and the effect of multiple scatterings becomes more significant.

It should be noted, however, that a DM particle cannnot be scattered an unlimited number of times. Once a DM particle has received enough energy to escape from the galaxy after multiple scatterings, further scatterings do not contribute to the constraint. Therefore, for a given $m_\chi$, each DM particle has a maximum number of scatterings, which is dependent on both the formation history of the DM halo and the interaction kinematics. These are beyond the scope of the present paper and we present the single scattering constraints as our main results.

\subsection{Comparison with other constraints}

\begin{figure}[htbp]
    \centering
    \includegraphics[width=1\linewidth]{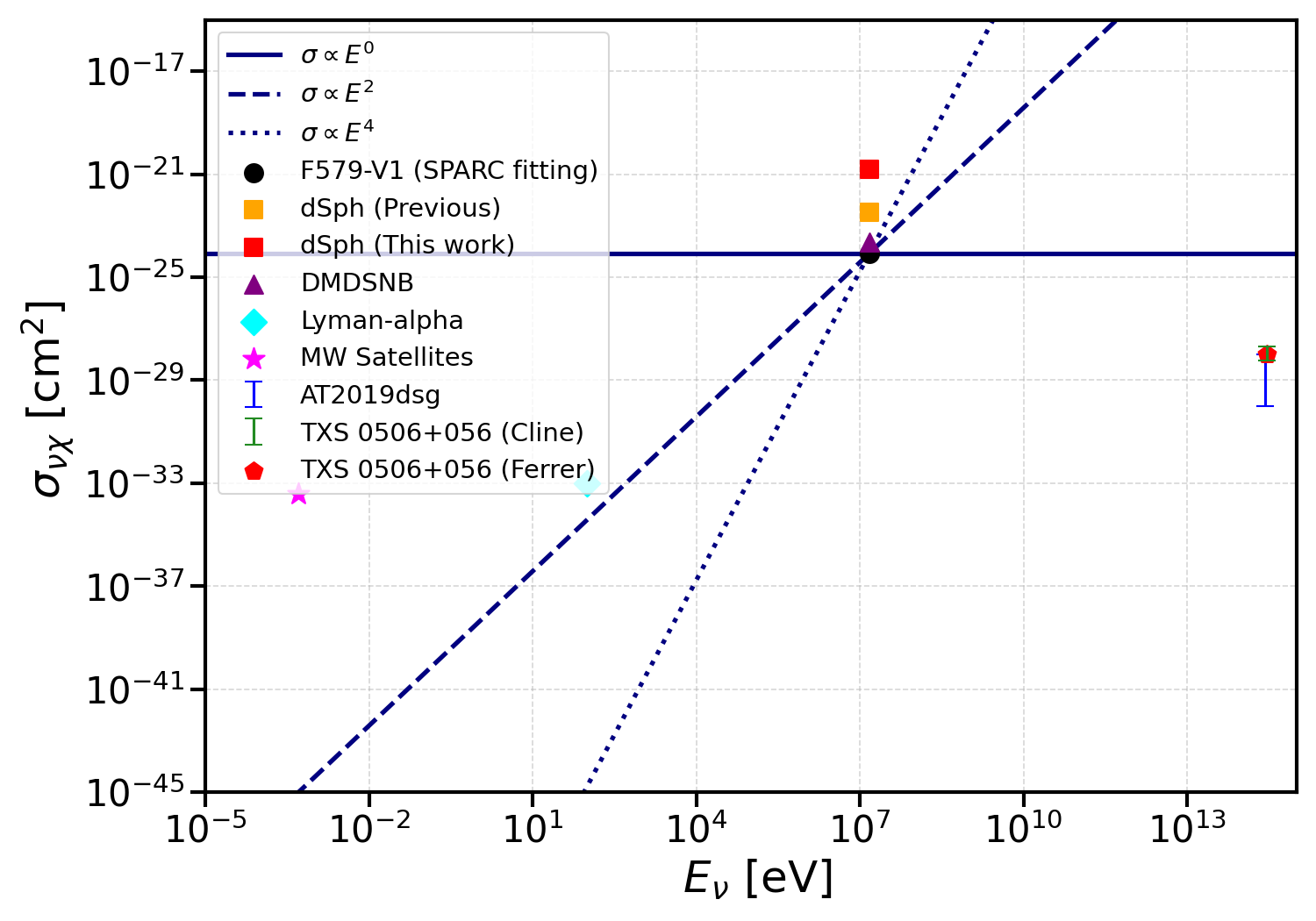}
    \caption{Comparison of upper limits on the dark matter-neutrino scattering cross section at a fixed dark matter mass of $m_\chi=1\, \rm GeV$. The black dot represent our calculated constraint from F579-V1 with a power-law energy dependence with $n=0,2,4$. For constraints using CCSN neutrinos, the orange square, the red square and purple triangle show the previous Fornax dSph limit \cite{Heston:2024ljf}, Fornax limit from this work, and the DMDSNB bound \cite{Chauhan:2025hoz}, respectively. Cosmological bounds from the Lyman-$\alpha$ forest \cite{Wilkinson:2014ksa} (cyan diamond) and Milky Way satellites \cite{Akita:2023yga} (magenta star) are shown for the constant cross section model ($n=0$). Constraints from AT2019dsg \cite{Fujiwara:2023lsv} (blue bar) and TXS 0506+056 (green bar for Cline et al.~\cite{Cline:2022qld} and red pentagon for Ferrer et al.~\cite{Ferrer:2022kei}) are evaluated assuming no dark matter annihilation. For AT2019dsg, the upper and lower ends of the error bar correspond to dark matter spike slopes of $\gamma_{\rm sp}=1.5$ and 2.25, respectively. For TXS 0506+056, the green bar reflects $\gamma_{\rm sp}=3/2$ and 7/3, while the red pentagon adopts $\gamma_{\rm sp}=7/3$ with a neutrino emission radius in the blazar jet of $R_{\rm em}\sim R_{\rm BLR}.$}
    \label{sigma_comparison}
\end{figure}

Figure \ref{sigma_comparison} compares our derived upper limits on the dark matter-neutrino scattering cross section using F579-V1 LSBG (black dot) with various existing constraints at a fixed dark matter mass of $m_\chi = 1\,\rm GeV$. For CCSN neutrinos with a typical energy of $E_\nu \sim 15\, \rm MeV$, we compare our result with the previous limit from the Fornax dSph and the bound from the diffuse supernova neutrino \cite{Chauhan:2025hoz} (DMDSNB, purple triangle). For Fornax, we show both the limit derived in \cite{Heston:2024ljf} (orange square) and that recomputed with our modified method (red square). Our LSBG constraint is about three orders of magnitude stronger than the Fornax limit derived with the same method, and is slightly stronger than the DMDSNB bound. Furthermore, we extend our comparison to low-energy cosmological bounds and high-energy neutrino events by adopting energy-dependent cross section models $\sigma_{\nu \chi}\propto E_\nu^n \, (n=0,2,4)$. The cosmological constraints from the Lyman-$\alpha$ \cite{Wilkinson:2014ksa} (cyan diamond) and Milky Way satellites \cite{Akita:2023yga} (magenta star) show strong bound within the constant cross section model. Constraints from high-energy events, the 270 TeV tidal disruption event 2019dsg \cite{Fujiwara:2023lsv} (blue bar) and the 290 TeV blazar flare TXS0506+056 \cite{Cline:2022qld,Ferrer:2022kei} (green bar and red pentagon), are compared assuming models with no dark matter annihilation. Across all energy dependencies, these high-energy constraints remain more stringent than our LSBG limit.

\section{Conclusion} \label{conclusion}
In this work, we derived an upper bound on the interaction cross section between dark matter particles and neutrinos by considering dark matter core formation driven by CCSN neutrinos in Low Surface Brightness Galaxies (LSBGs). We extended the framework of the previous study on dSph \cite{Heston:2024ljf} to LSBGs, which possess massive dark matter halos and abundant stellar populations, as well as the modification to the counting of scatterings. Utilizing the kinematic data and core radius estimations analyzed by the de Blok, DensePak, and SPARC survays, we calculated the stellar mass and the neutrino energy emitted by past CCSNe for each galaxy. We then estimated the fraction of energy required to transform an initial NFW cuspy profile into the observed core. Under the assumption that the $\nu-$DM interactions from CCSNe are the only source of core formation, we calculated the upper limits on the scattering cross section. 

As a result, for CCSN neutrinos with energies of $\mathcal O(10)\,\rm MeV$, we found that F579-V1 provides the most stringent limit in the low-mass regime ($m_\chi \leq 35\,\rm GeV$), while UGC06628 yields the strongest constraint in the high-mass regime ($m_\chi > 35\,\rm GeV$). Compared with the recalculated limits for Fornax dSph, our results show a significant improvement by approximately three orders of magnitude in the constant cross section regime, whereas compared with DMDSNB limits, it indicates a slightly stronger constraints.

There are several uncertainities in our adopted framework that are worth mentioning. To calculate the total neutrino energy emitted from past CCSNe, we assumed a specific initial mass function. 
For instance, varying the IMF slope above $8M_\odot$, which can be considered to have the largest contribution, by $\pm 0.3$ from $2.3$
\cite{Kroupa:2002ky} results in a $-45\%$ to $+59\%$ change in $N_{\rm CCSNe}$. This translates into an uncertainty in $\sigma_{\nu\chi}$ within a factor of two. Furthermore, we utilized a flavor-independent and time-integrated neutrino energy spectrum. In reality, the energy spectrum of CCSN neutrinos differ between flavors, which would yield different limits for each flavor. In addition, since the DM column density is highly concentrated around the center, we assumed that he interactions occur within the central cusp region, neglecting the CCSNe spatial distribution, and that each scattering transfers the maximum energy.

However, there are several points that make our limits conservative. First, the energy required for dark matter core formation was evaluated using the upper limit of the observed core radius $r_c+1\sigma$, which enables us to calculate the constraints conservatively. Second, in deriving the constraints in energy injection regime, each DM particle is assumed to be scattered at most once; allowing multiple scatterings would strengthen the constraints. Moreover, in our framework, we assumed that $\nu-$DM interactions are the only source of dark matter core formation, neglecting other astrophysical effects. For instance, baryonic feedback from supernovae is also expected to contribute to core formation \cite{Mashchenko:2007jp, Governato:2009bg, Pontzen:2011ty, DiCintio:2013qxa, Chan:2015tna, Dutton:2016okm, Tollet:2015gqa, Freundlich:2019lkx, Burger:2021sep}. If such effects were included in our calculation, the required energy from the $\nu$-DM interactions would be reduced, which would yield even stronger constraints on the cross section.

\section{acknowledgements}
We thank John Beacom, Takumi Iwahori, and Satoshi Shirai for useful comments. This research is supported in part by JSPS KAKENHI grant No.~23H04899 as well as by the World Premier International Research Center Initiative (WPI Initiative), MEXT, Japan.

Software: NumPy\cite{Harris:2020xlr}, SciPy\cite{Virtanen:2019joe}, Matplotlib\cite{Hunter:2007ouj}, pandas\cite{McKinney:2010nts}

\bibliography{main}

\end{document}